\documentclass[11pt,dvips]{article}

\usepackage{epsfig,times} 
\usepackage{floatflt}
\makeatletter
\renewcommand{\section}{\@startsection{section}{1}{0in}
	{0.4\baselineskip}{0.1\baselineskip}{\Large\bf}}
\renewcommand{\subsection}{\@startsection{subsection}{2}{0in}
	{0.25\baselineskip}{-\baselineskip}{\large\bf}}
\renewcommand{\subsubsection}{\@startsection{subsubsection}{3}{0in}
	{0.1\baselineskip}{-\baselineskip}{\normalsize\bf}}
\makeatother
\newcommand{\Dnet}{\ensuremath{D_{\rm net}}}
\newcommand{\Dfg}{\ensuremath{D_{\rm fg}}}
\newcommand{\Dbg}{\ensuremath{D_{\rm bg}}}
\begin{document}
%
\makeatletter\newcommand{\ps@icrc}{
\renewcommand{\@oddhead}{\slshape{HE 6.4.13}\hfil}}
\makeatother\thispagestyle{icrc}
%
\markright{HE 6.4.13}
\begin{center}
%
{\LARGE \bf Emulsion Chamber Densitometry by Macroscopic Digital Imaging}
\end{center}

\begin{center}
%
%
{\bf E. L. Zager, R. J. Wilkes, J. J. Lord}\\
{\it Department of Physics, University of Washington, Seattle WA 98195, USA}
\end{center}

\begin{center}
{\large \bf Abstract\\}
\end{center}
\vspace{-0.5ex}
%
%
%
Spot density is commonly used as an indication of shower energy in
emulsion chambers.  In a system originally developed for JACEE
analysis, the optical density of a spot on x-ray film is
estimated from macroscopic digital images.  The spot's size is used to
compensate for the lack of dynamic range obtainable with digital
imaging hardware.  These densities are compared to manually measured
densities.

%
\vspace{1ex}

%
%
\section{Introduction}
Previous work described a computerized system for automatically
reconstructing cascade tracks from digital images of an emulsion
chamber's x-ray films, originally developed for JACEE analysis (Zager,
1997).  We build on that work to describe how spot densities may be
measured from information available to the computer as a byproduct of
the track reconstruction process.

The result of the reconstruction process is a list of tracks in the
chamber, and the spots which make up those tracks.  The next step in
our analysis is to measure the density on the x-ray films by hand for
each spot.  Since the computer has both the
images of those films and the location of each spot, it is reasonable
to try to automate this process.

Spot density is commonly used as an indication of shower energy in
emulsion chambers (Burnett, 1986).  Traditionally, spot density is
measured by an optical instrument which measures the trasmission of
light through a 200--300 micron slit.  Recently this technique has
been extended to micro-densitometry, in which an optical density
measurement is made by computer analysis of a microscopic image of
that film.  Here we describe attempts to measure density by
computer analysis of an image of an entire x-ray film.  This image is
necessarily of far lower resolution than the microscopic image used for
micro-densitometry, so a new technique must be developed.

\section{Technique}
We use optical density to estimate $N_e$, the number of electrons in a
shower.  $N_e$ is directly related to the total energy in the
electromagnetic coomponent of a shower.  The relation between $N_e$
and density is dependent on film and development conditions, so for
each set of x-ray films, a calibration is done between the density of
a given spot and the number of singly-ionizing tracks seen under a
microscoope in the nuclear emulsion.  Optical density is defined as
\begin{equation}
D = -\log_{10}\left(\frac{I_{\it transmitted}}{I_{\it incident}}\right).
\end{equation}
To generate a quantity which is independent of $I_{\it incident}$, we
use 
\begin{equation}
\Dnet \equiv \Dfg - \Dbg = -\log_{10}\left(\frac{I_{\it fg}}{I_{\it bg}}\right),
\end{equation}
where \Dfg\ and \Dbg\ are the densities of the foreground (the spot) and 
the background (the neighborhood of the spot), respectively.

The x-ray film is imaged by a CCD camera at approximately
$3400\times2700$ pixels, 12 bit grayscale.  The
film itself is $50\times40$ cm, so a pixel is approximately 150 microns 
on an edge.  

\subsection{Measuring the Background Density}

The background density of the film itself varies due to slight
irregularities in the development process.  The digital image of that
film has further variations in background density due to illumination
of the film and the optical qualities of the lens used.  These
combine to produce significant variations in the local background
density, so it is necessary to measure \Dbg\ in the neighborhood of
each spot.

To rephrase the problem, we wish to measure the average density over
all pixels in the neighborhood of a spot which are not part of that
spot or of any other spot.  Fortunately in the process of identifying
spots, the program has already identified every pixel of the image
which belong to a spot.  Since a spot has blurry edges, it is
important to exclude the edges from the calculation.  

To exclude these edges we use a standard image processing technique
called dilation, which causes the edges of a feature to grow by
one pixel (Russ, 1995).  By appling fourteen dilation rounds, we move
the edge of each spot out fourteen pixels, or about 2 mm.  This is
sufficient to exclude the edges of all but the largest showers.  We
can then safely consider all the remaining pixels in the neighborhood
to be background, unaffected by showers.

\subsection{Foreground Density}

The size of a single pixel in our image is comparable to the size of
the slit in optical densitometry.  Since a CCD has a linear response
to intensity, it should be possible to measure the density by taking
the darkest pixel of the spot as $I_{\it fg},$ and the average
intensity of the background near the spot as $I_{\it bg}.$
%
%

\begin{floatingfigure}{4in}
\epsfig{file=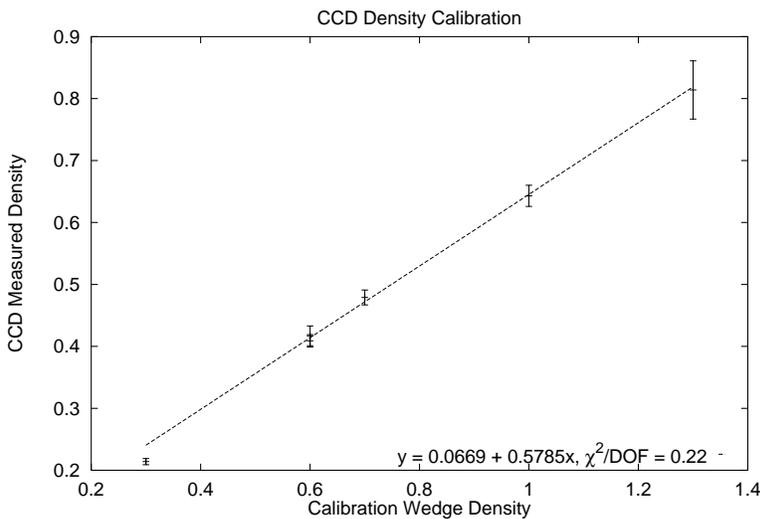,width=3.95in}
\caption{Calibration of CCD Density against Neutral-Density Filters}
\label{density-ccd}
\end{floatingfigure}

Figure \ref{density-ccd} shows the calibration of our CCD and image
acquisition system against Bausch \& Lomb neutral density filters.  Our
system is linear, although it reads systematically low by a scale
factor of about 1.7.  We have seen similar scale factors when
cross-calibrating other densitometers, so this is not a cause for
concern (Olson, 1995).

With a 12-bit grayscale image, the\\ largest \Dnet possible is \\
$\Dnet = -\log_{10}(1/4096) = 3.6.$\\ However, under realistic
illumination conditions we find a situation where $I_{\rm fg}$ is
about 75\% of the full range, and $I_{\rm bg}$ is about 10\%.  This
translates to a maximum \Dnet\ of about 0.9.  The density of spots on
an x-ray film varies with the development of the film: the longer the
development, the greater the density of a given spot.  We tend to
develop our films to produce a maximum \Dnet\ around 2.0.  This leaves
the bulk of spots well under
\Dnet\ of 1.0.  The simple method described above may be adequate for 
the majority of events, but we need a different method to estimate the
density of the highest energy events.

\subsection{Spot Area}
Although density may saturate for higher energy events, a spot's size
will continue to grow.  Shower energy has been successfully related to
the area of a microscopic emulsion image (Fuki, 1995).  The concept
here is similar, but at a macroscopic scale.  To turn spot area into a
useful measurement, we need two things: a way to accurately measure
the area, and a correlation between area and density.

Measuring the area of a spot is somewhat tricky.  The lateral profile
of charged particles in a cascade falls off approximately as $r^{-1}$
(Olson, 1995), so the spot has very soft edges and blurs seamlessly
into the background.  Since there is no hard edge, we somewhat
arbitrarily choose one which we can construct.  We compute the spot's
area as the sum of all pixels contiguous with the spot which are
darker than a given threshold.  The selection of this threshold is
discussed below.  Once again, this information is a byproduct of the
track reconstruction phase.

\subsection{Adjustment for Inclination}
A simple model of an inclined event would indicate that the area for
an inclined shower varies as $[cos\theta]^{-1}$ due to the projection
of the spot onto the film plane. But density also has a slope
dependence due to physical characteristics of the x-ray film.
Emulsion thickness, grain size, and the presence of a second emulsion
on the back of the film all contribute.  A study of the effect of
inclination on density found results consistent with a simple
$[\cos\theta]^{-1}$ scaling, but could not rule out exponents in the
range (-0.8) -- (-1.2) (Olson, 1995).  We stick to a simple model in
which the effect of inclination on spot area will tend to be canceled
out by the effect on density, so inclination is neglected.

\section{Results}
%
%
We applied the techniques above to a set of x-ray films exposed during
the thirteen day JACEE-13 Antarctic balloon flight (Wilkes, 1995).  Tracks
in one emulsion chamber were reconstructed manually, then densities
measured manually by micro-densitometry.  We used the software to
generate an independent map and set of spot measurements.  Finally we
matched the two sets of tracks.

%
%
\subsection{Density-Density Correlation}

\begin{figure}
\label{density-density}
\centering
\epsfig{file=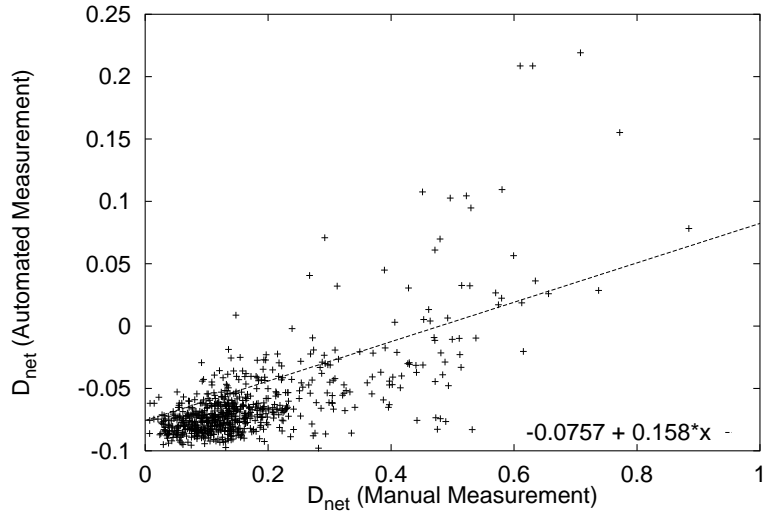,width=4in}
\caption{\Dnet\ measured by \Dfg\ technique compared to \Dnet\ measured
by manually with micro-densitometry.  Variation is greater past
density 1.0. Negative values are an artifact of the CCD density
calibration.}
\end{figure}

\Dnet\ measured manually and \Dnet\ measured by
the software are loosely correlated, as shown in figure
\ref{density-density}.  At the low end, $\Dnet < 0.2,$ the errors are
probably dominated by the measurement of \Dfg. When measuring \Dfg\
manually, the image is carefully aligned so that the darkest part of
the spot is centered in the window.  The automatic system does not
have this luxury.  These low-density spots are small, generally 7--8
pixels on an edge.  The darkest point may or may not align with the
center of the pixel.  Large spots are less vulnerable to this problem
because the darkest part of the spot will occupy more than one pixel.
%
%
\subsection{Density-Area Correlation}
\begin{figure}
\centering
\epsfig{file=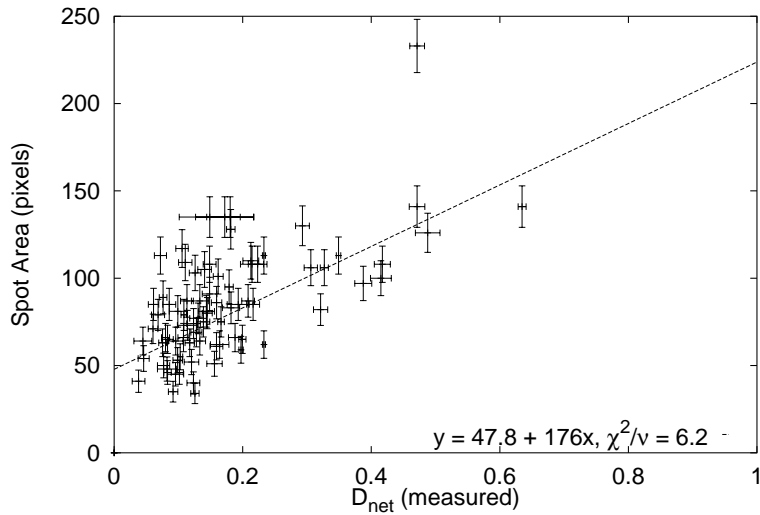,width=4in}
\caption{Spot area compared to \Dnet\ measured
by manual approach with micro-densitometry.  Shown are all spots
measured by both systems on film ZZ-C16C. Variation is greater past
density 1.0.}
\label{area-density}
\end{figure}

Figure \ref{area-density} shows the correlation between spot area
and density.  Only measurements from one x-ray film are shown because
the area of a spot is dependent in part on the threshold intensity
chosen to separate an image from its background.  As the track
reconstruction software adjusts this threshold differently for each
image, it is only meaningful to compare spot areas measured on the
same image.

Clearly we would like to be able to compare density, and therefore
area, across different images.  This requires a consistent threshold
for each image.  Another pass through the images can easily accomplish
this once track reconstruction is complete.

The correlation between area and density looks good above \Dnet\ =
0.3.  The variation seen at the lower end is probably due to the
difficulty is discriminating between pixels which make up the image,
and pixels which make up the background.  Setting the threshold
between foreground and background to be higher may produce better
results, but may also tend to produce precision errors.  These
smaller spots are images of as few as 36 pixels.  Increasing the
discrimination threshold will reduce these spots further.
\section{Conclusions}
A careful study of the discrimation between the spot and its
background is likely to benefit both the measurement of area and of
literal density.  The measurement of density may also be helped by
fitting the measured intensities to an assumed lateral distribution
function.  By combining the literal density measurement with
measurement of spot area, the automatic system may be able to produce
good measurements of spot density.  It is likely that this combination
would rely more on density for lower-energy particles, and more on
area for higher energy particles.  This would allow us to estimate
shower energy very quickly compared to current methods.  The technique
is not limited to analysis of x-ray films.  Real-time electronic
detectors which produce similar images of a cascade could employ the
same method.

The author wishes to acknowledge the densitometry work of Bjorn S. Nilsen.

%
%
%
%
\section{References}
%
%
Burnett, T. H. {\it et al.}, 1986 Nucl Inst Meth, A251:583-595.\\
Fuki, M. and Takahashi, Y. Proc. 25th ICRC (Rome, 1995) 3, 677.\\
Olson, E. D. 1995. The Energy Spectrum of Primary Cosmic Ray Nuclei
above 1 TeV per Nucleon. Ph.D. thesis, Universion of Washington,
Seattle.\\
Russ, John C. 1995, {\em The Image Processing Handbook.} CRC Press, Boca Raton.
Wilkes, R. J. {\it et al}, Proc. 24th ICRC (Rome, 1995) 3, 615.\\
Zager, E. L. {\it et al}, Proc. 25th ICRC (Durban, 1997) 7, 293.\\
\end{document}